\documentclass[12pt]{article}
\renewcommand{\thefootnote}{\fnsymbol{footnote}}

\begin{document}

\baselineskip 6mm
\begin{flushright}
{\tt TMI-00-2 \\
     December 2000}
\end{flushright}
\thispagestyle{empty}
\vspace*{5mm}
\begin{center}
   {\Large {\bf Scalar partner of $Z^{0}$-boson with larger 
                mass value than $Z^{0}$-boson mass in 
                Subquark model}}
\vspace*{20mm} \\
  {\sc Takeo~~Matsushima}
 \footnote[1]
  {
   $\dagger$ e-mail : mtakeo@eken.phys.nagoya-u.ac.jp
 }
\vspace{5mm} \\
  \sl{1-17-1 Azura Ichinomiya,$^\dagger$}\\
  \it{Aichi-Prefecture, Japan}
\end{center}

\vspace{15mm}  

\begin{center}
{\bf Abstract} \\
\vspace{3mm}
\begin{minipage}[t]{120mm}
\baselineskip 5mm
{\small
  The subquark model previously proposed by us showes that 
  the intermediate $Z^{0}$-boson is realized as the composite
  object and its scalar partner has the mass value larger 
  than $Z^{0}$-boson mass, which is about $110$ GeV}.
\end{minipage}

\end{center}


\newpage
\baselineskip 18pt
\section{Introduction}
\hspace*{\parindent}
  The discovery of the top-quark[1] has finally confirmed the 
 existence of three quark-lepton symmetric generations.
 So far the standard $SU(2)_{L}\otimes{U(1)}$ model
 (denoted by SM)
 has successfully explained various experimental evidences.
 Nevertheless, as is well known, the SM is not regarded as
 the final theory because it has many arbitrary parameters, e.g.,
 quark and lepton masses, quark-mixing parameters and the Weinberg 
 angle, etc. . Therefore it is meaningful to investigate the
 origins of these parameters and the relationship among them.
  In order to overcome such problems some attempts have done, e.g.,
 Grand Unification Theory (GUT); Supersymmetry(SUSY); SUGY-GUT;
  Composite model; etc. .
 In the GUT scenario quarks and leptons are elementary fields
 in general. On the contrary in the composite scenario they are
 literally the composite objects constructed from the elementary
 fields (so called ``preon''). The lists of various related 
 works are in ref.[2]. If quarks and leptons are 
 elementary, in order to solve the above problems it is necessary 
 to introduce some external relationship or symmetries among them. 
  On the other hand the composite models have ability to explain
 the origin of these parameters in terms of the substructure 
 dynamics of quarks and leptons. Further, the composite scenario 
 naturally leads us to the thought that the intermediate vector bosons
 of weak interactions (${\bf W,Z}$) are not elementary gauge fields
 (which is so in the SM)
 but composite objects constructed from preons (same as ${\bf \rho}$
 -meson from quarks). Many studies based on such conception have 
 done after Bjorken's[3] and Hung and Sakurai's[4] suggestions of the 
 alternative way to unified weak-electromagnetic gauge theory[5-11].
 In this scheme the weak interactions are regarded as the effective
 residual interactions among preons. The fundamental fields for
 intermediate forces are massless gauge fields belonging to
 some gauge groups and they confine preons into singlet states
 to build quarks and leptons and ${\bf W,Z}$.
\par
  The conception of our model is that the fundamental interacting
 forces are all originated from massless gauge fields belonging
 to the adjoint representations of some gauge groups which have 
 nothing to do with the spontaneous breakdown and that the 
 elementary matter fields are only one kind of spin-1/2 preon
 and spin-0 preon carrying common ``$e/6$'' electric charge ($e>0$).
 Quarks, leptons and ${\bf W,Z}$ are all composites of them and
 usual weak interactions are regarded as effective residual
 interactions. Based on such idea we consider the underlying gauge 
 theory in section 2 and composite model in section 3. In section
  4 we discuss about the mass(denoted by $M(S^{0})$) of the scalar
 partner(denoted by $S^{0}$) of $Z^{0}$-boson.
   
 \section{  Gauge theory inspiring quark-lepton composite scenario}
 \hspace*{\parindent}
 
  In our model the existence of fundamental matter fields (preon)
 are inspired by the gauge theory with Cartan connections[14]. 
 Let us briefly summarize the basic features of that. Generally 
 gauge fields, including gravity, are considered as geometrical  
 objects, that is, connection coefficients of principal fiber 
 bundles. It is said that there exist some different points 
 between Yang-Mills gauge theories and gravity, though both 
 theories commonly possess fiber bundle structures. The latter has 
 the fiber bundle related essentially to 4-dimensional space-time
 freedoms but the former is given, in an ad hoc way, the one with 
 the internal space which has nothing to do with the space-time 
 coordinates. In case of gravity it is usually considered that 
 there exist ten gauge fields, that is, six spin connection fields 
 in $SO(1,3)$ gauge group and four vierbein fields in $GL(4,R)$ 
 gauge group from which the metric tensor ${\bf g}^{{\mu}{\nu}}$ 
 is constructed in a bilinear function of them. Both altogether 
 belong to  Poincar\'e group $ISO(1,3)=SO(1,3)\otimes{R}^4$ 
 which is semi-direct product. In this scheme spin connection 
 fields and vierbein fields are independent but only if there is 
 no torsion, both come to have some relationship. Seeing this, 
 $ISO(1,3)$ gauge group theory has the logical weak point not to  
 answer how two kinds of gravity fields are related to each other 
 intrinsically.
\par   
In the theory of Differential Geometry, S.Kobayashi has investigated 
 the theory of ``Cartan connection''[15]. This theory, in fact, 
 has ability to reinforce the above weak point. The brief 
 recapitulation is as follows. Let $E(B_n,F,G,P)$ be a fiber bundle
 (which we call Cartan-type bundle) associated with a principal 
 fiber bundle $P(B_n,G)$ where $B_n$ is a base manifold with  
 dimension ``$n$'', $G$ is a structure group, $F$ is a fiber space 
 which is homogeneous and diffeomorphic with $G/G'$ where $G'$ 
 is a subgroup of $G$. Let $P'=P'(B_n,G')$ be a principal 
 fiber bundle, then $P'$ is a subbundle of $P$.  
 Here let it be possible to decompose the Lie algebra ${\bf g}$ of 
 $G$ into the subalgebra ${\bf g}'$ of $G'$ and a vector space 
 ${\bf f}$ such as : 
 \begin{equation}
     {\bf g}={\bf g}'+{\bf f},\hspace{1cm}{\bf g}'\cap{\bf f}=0,      \label{1}
 \end{equation}
 
 \begin{equation}
     [{\bf g'},{\bf g'}]\subset {\bf g'},                             \label{2}
 \end{equation}
 
 \begin{equation}
     [{\bf g'},{\bf f}]\subset {\bf f},                               \label{3}
 \end{equation}
 
 \begin{equation}
     [{\bf f},{\bf f}]\subset {\bf g'},                               \label{4}  
 \end{equation}
 where $dim{\bf f}=dimF=dimG-dimG'=dimB_n=n$. 
 The homogeneous space $F=G/G'$ is said to be ``weakly reductive'' 
 if there exists a vector space ${\bf f}$ satisfying Eq.(1) and (3).
 Further $F$ satisfying Eq(4) is called ``symmetric space''. 
 Let ${\bf \omega}$ denote the connection form of $P$ and 
 $\overline{\bf \omega}$ be the restriction of ${\bf \omega}$ to
 $P'$. Then $\overline{\bf \omega}$ is a ${\bf g}$-valued linear
 differential 1-form and we have :
 \begin{equation}
     {\bf \omega}=g^{-1}\overline{\bf \omega}g+g^{-1}dg,              \label{5}
 \end{equation}
 where $g\in{G}$, $dg\in{T_g(G)}$. ${\bf \omega}$ is called the  
 form of ``Cartan connection'' in $P$. 
\par
 Let the homogeneous space $F=G/G'$ be weakly reductive. The 
 tangent space $T_O(F)$ at $o\in{F}$ is isomorphic with ${\bf f}$
 and then $T_O(F)$ can be identified with ${\bf f}$ and also 
 there exists a linear ${\bf f}$-valued differential 
 1-form(denoted by ${\bf \theta}$) which we call the 
 ``form of soldering''. Let ${\bf \omega}'$ 
 denote a ${\bf g}'$-valued 1-form in $P'$, we have :
 \begin{equation}
    \overline{\bf \omega}={\bf \omega}'+{\bf \theta}.                  \label{6}
 \end{equation}
 The dimension of vector space ${\bf f}$ and the dimension of  
 base manifold $B_n$ is the same ``$n$'', and then ${\bf f}$  
 can be identified with the tangent space of $B_n$ at the same 
 point in $B_n$ and ${\bf \theta}$s work as $n$-bein fields. In this 
 case  ${\bf \omega}'$ and ${\bf \theta}$ unifyingly belong to 
 group $G$. Here let us call such a mechanism ``Soldering Mechanism''.
\par
 Drechsler has found out the useful aspects of this theory and 
 investigated a gravitational gauge theory based on 
 the concept of the Cartan-type bundle equipped with the 
 Soldering Mechanism[16]. He considered $F=SO(1,4)/SO(1,3)$ 
 model. Homogeneous space $F$ with $dim=4$ solders 4-dimensional 
 real space-time. The Lie algebra of $SO(1,4)$ corresponds to 
 ${\bf g}$ in Eq.(1), that of $SO(1,3)$ corresponds to ${\bf g}'$
 and ${\bf f}$ is 4-dimensional vector space. The 6-dimensional 
 spin connection fields are ${\bf g}'$-valued objects and vierbein 
 fields are ${\bf f}$-valued, both of which are unified into 
 the members of $SO(1,4)$ gauge group. We can make the metric 
 tensor ${\bf g}^{{\mu}{\nu}}$ as a bilinear function of 
 ${\bf f}$-valued vierbein fields. Inheriting Drechsler's study 
 the author has investigated the quantum theory of gravity[14]. 
 The key point for this purpose is that $F$ is a symmetric 
 space because ${\bf f}$s are satisfied with Eq.(4). 
 Using this symmetric nature we can pursue making a 
 quantum gauge theory, that is, constructing ${\bf g}'$-valued 
 Faddeev-Popov ghost, anti-ghost, gauge fixing, gaugeon 
 and its pair field as composite fusion fields of 
 ${\bf f}$-valued gauge fields by use 
 of Eq.(4) and also naturally inducing BRS-invariance. 
\par
 Comparing such a scheme of gravity, let us consider 
 Yang-Mills gauge theories. Usually when we make the Lagrangian 
 density ${\cal L}=tr({\cal F}\wedge{\cal F}^{\ast})$ 
 (${\cal F}$ is a field strength), we must 
 borrow a metric tensor ${\bf g}^{{\mu}{\nu}}$ from gravity to 
 get ${\cal F}^{\ast}$ and also for Yang-Mills gauge fields to 
 propagate in the 4-dimensional real space-time. This seems to 
 mean that ``there is a hierarchy between gravity and other three 
 gauge fields (electromagnetic, strong, and weak)''. But is it   
 really the case ? As an alternative thought we can think that  
 all kinds of gauge fields are ``equal''. Then it would be natural 
 for the question ``What kind of equality is that ?'' to arise. 
 In other words, it is the question that ``What is the minimum 
 structure of the gauge mechanism which four kinds of forces 
 are commonly equipped with ?''. For answering this question, 
 let us make a assumption : ``\begin{em}Gauge fields 
 are Cartan connections equipped with Soldering 
 Mechanism\end{em}.'' In this meaning all 
 gauge fields are equal. If it is the case three gauge fields 
 except gravity are also able to have their own metric 
 tensors and to propagate in the real space-time without 
 the help of gravity. Such a model has already investigated 
 in ref.[14]. 
\par
 Let us discuss them briefly. It is found that there 
 are four types of sets of classical groups with small dimensions 
 which admit Eq.(1,2,3,4), that is, $F=SO(1,4)/SO(1,3)$, 
 $SU(3)/U(2)$, $SL(2,C)/GL(1,C)$ and $SO(5)/SO(4)$ with 
 $dimF=4$[17]. Note that the quality of ``$dim\hspace{1mm}4$''
  is very important because it guarantees $F$ to solder 
 to 4-dimensional real space-time and all gauge fields 
 to work in it. The model of $F=SO(1,4)/SO(1,3)$ for gravity 
 is already mentioned. 
 Concerning other gauge fields, it seems to be appropriate  
 to assign $F=SU(3)/U(2)$ to QCD gauge fields, 
 $F=SL(2,C)/GL(1,C)$ to QED gauge fields and 
 $F=SO(5)/SO(4)$ to weak interacting gauge fields. Some 
 discussions concerned are following. In general, matter 
 fields couple to ${\bf g}'$-valued gauge fields. As for   
 QCD, matter fields couple to the gauge fields of $U(2)$
 subgroup but $SU(3)$ contains, as is well known, three types 
 of $SU(2)$ subgroups and then after all they couple 
 to all members of $SU(3)$ gauge fields. In case of QED,  
 $GL(1,C)$ is locally isomorphic with $C^1\cong{U(1)}\otimes{R}$. 
 Then usual Abelian gauge fields are assigned to $U(1)$ 
 subgroup of $GL(1,C)$. Georgi and Glashow suggested that  
 the reason why the electric charge is quantized comes from 
 the fact that $U(1)$ electromagnetic gauge group is 
 a unfactorized subgroup of $SU(5)$[18]. Our model is 
 in the same situation because $GL(1,C)$ a unfactorized 
 subgroup of $SL(2,C)$. For usual electromagnetic $U(1)$ 
 gauge group, the electric charge unit ``$e$''$(e>0)$ is for   
 $one\hspace{2mm} generator$ of $U(1)$ but in case of $SL(2,C)$ 
 which has $six\hspace{2mm} generators$, the minimal 
 unit of electric charge 
 shared per one generator must be ``$e/6$''. This suggests 
 that quarks and leptons might have the substructure 
 simply because $e,\hspace{1mm}2e/3,\hspace{1mm}e/3>e/6$.
  Finally as for weak interactions we adopt 
 $F=SO(5)/SO(4)$. It is well known that $SO(4)$ is 
 locally isomorphic with $SU(2)\otimes{SU(2)}$. Therefore 
 it is reasonable to think it the left-right symmetric 
 gauge group : $SU(2)_L\otimes{SU(2)}_R$. As two $SU(2)$s are 
 direct product, it is able to have coupling constants 
 (${\bf g}_L,{\bf g}_R$) independently. 
 This is convenient to explain the fact of the disappearance 
 of right-handed weak interactions in the low-energy region. 
 Possibility of composite structure of quarks and leptons 
 suggested by above $SL(2,C)$-QED would introduce 
 the thought that the usual left-handed weak interactions 
 are intermediated by massive composite vector bosons as 
 ${\bf \rho}$-meson in QCD and that they are residual 
 interactions due to substructure dynamics of quarks 
 and leptons. The elementary massless gauge fields ,``\begin{em}
 as connection fields''\end{em}, relate intrnsically to the
  structure of the real space-time manifold but on the other hand 
 the composite vector bosons have nothing to do with it. 
 Considering these discussions, we set the assumption : 
 ``\begin{em}All kinds of gauge fields are elementary massless 
 fields, belonging to spontaneously unbroken 
 $SU(3)_C\otimes{SU(2)}_L\otimes{SU(2)}_R\otimes{U(1)}_{e.m}$ 
 gauge group and quarks and leptons and {\bf W, Z} are all 
 composite objects of the elementary matter fields\end{em}.''

\section{Composite model}
 \hspace*{\parindent}
 Our direct motivation towards compositeness of quarks and leptons 
 is one of the results of the arguments in Sect.2, that is, 
 $e,\hspace{1mm}2e/3,\hspace{1mm}e/3>e/6$. 
 However, other several phenomenological 
 facts tempt us to consider a composite model, e.g.,  
 repetition of generations, quark-lepton parallelism of weak 
 isospin doublet structure, quark-flavor-mixings, etc..
 Especially Bjorken[3]'s and Hung and Sakurai[4]'s suggestion 
 of an alternative to unified weak-electromagnetic gauge theories
 have invoked many studies of composite models including 
 composite weak bosons[5-11]. Our model is in the line of   
 those studies. There are two ways to make composite 
 models, that is, ``Preons are all fermions.'' or ``Preons are   
  both fermions and bosons (denoted by FB-model).'' 
 The merit of the former is that it can avoid the problem 
 of a quadratically divergent self-mass of elementary scalar 
 fields. However, even in the latter case such a disease 
 is overcome if both fermions and bosons are the 
 supersymmetric pairs, both of which carry the same quantum 
 numbers except the nature of Lorentz transformation (
 spin-1/2 or spin-0)[19]. Pati and Salam have suggested    
 that the construction of a neutral composite object 
 (neutrino in practice) needs both kinds of preons, fermionic 
 as well as bosonic, if they carry the same charge for the 
 Abelian gauge or belong to the same 
 (fundamental) representation for the non-Abelian gauge[20].  
 This is a very attractive idea for constructing the minimal 
 model. Further, according to the representation theory of  
 Poincar\'e group both integer and half-integer 
 spin angular momentum occur equally for massless particles[21], 
 and then if nature chooses ``fermionic monism'', 
 there must exist the additional special reason to select it. 
 Therefore in this point also, the thought of the FB-model 
 is minimal. Based on such considerations we propose 
 a FB-model of \begin{em}``only one kind of spin-1/2 
 elementary field 
 (denoted by $\Lambda$) and of spin-0 elementary field
 (denoted by $\Theta$)''\end{em} (preliminary 
 version of this model has appeared in Ref.[14]). 
 Both have the same electric charge of ``$e/6$''  
 (Maki has first proposed the FB-model with the minimal 
 electric charge $e/6$. \cite{22})
\renewcommand{\thefootnote}{\arabic{footnote}}
\footnote{The notations of $\Lambda$ and $\Theta$ are 
 inherited from those in Ref.[22]. After this we 
 call $\Lambda$ and $\Theta$ ``Primon'' named by Maki 
 which means ``primordial particle''[22].}
 and the same transformation properties of the 
 fundamental representation ( 3, 2, 2) under the spontaneously unbroken   
 gauge symmetry of $SU(3)_C\otimes{SU(2)_L}\otimes{SU(2)_R}$
 (let us call $SU(2)_L\otimes{SU(2)_R}$ ``hypercolor gauge symmetry''). 
 Then $\Lambda$ and $\Theta$ come into the supersymmetric pair 
 which guarantees 'tHooft's naturalness condition[23]. 
 The $SU(3)_C$, $SU(2)_L$ and $SU(2)_R$ gauge fields 
 cause the confining forces with confining energy scales of 
 $\Lambda_c<< \Lambda_L<(or \cong) \Lambda_R$ (Schrempp and Schrempp 
 discussed this issue elaborately in Ref.[11]). 
 Here we call positive-charged 
 primons ($\Lambda$, $\Theta$) ``$matter$'' and negative-charged 
 primons ($\overline\Lambda$, $\overline\Theta$) ``$anti
 matter$''. Our final goal is to build quarks, leptons 
 and ${\bf W, Z}$ from $\Lambda$ ($\overline\Lambda$)
  and $\Theta$ ($\overline\Theta$).   
 Let us discuss that scenario next.
 \par
 At the very early stage of the development of the universe, 
 the matter fields ($\Lambda$, $\Theta$) and their antimatter fields 
 ($\overline{\Lambda}$, $\overline{\Theta}$) must have broken out  
 from the vaccum. After that they would have combined with each  
 other as the universe was expanding. That would be the 
 first step of the existence of composite matters. 
 There are ten types of them : 
\newline
 \hspace*{2mm}$spin\displaystyle{\frac{1}{2}}
 $\hspace{2.4cm}$spin\hspace{1mm}0$
 \hspace{2.1cm}$e.m.charge$
 \hspace{1.2cm}$Y.M.representation$\hspace{1cm}
{
 \setcounter{enumi}{\value{equation}}
 \addtocounter{enumi}{1}
 \setcounter{equation}{0} 
 \renewcommand{\theequation}{\theenumi\alph{equation}}
 \begin{eqnarray}
     \Lambda\Theta\hspace{2.5cm}\Lambda\Lambda,
     \Theta\Theta\hspace{3.1cm}
     \frac{1}{3}e\hspace{1.6cm}(\overline{3},1,1)\hspace{2mm}
     (\overline{3},3,1)\hspace{2mm}(\overline{3},1,3),\hspace{5mm}\\
     \Lambda\overline\Theta,\overline\Lambda\Theta\hspace{2cm}
     \Lambda\overline\Lambda,\Theta\overline\Theta\hspace{3.2cm}
     0\hspace{1.8cm}(1,1,1)\hspace{2mm}(1,3,1)\hspace{2mm}(1,1,3),\hspace{5mm}\\
     \overline\Lambda\overline\Theta\hspace{2.5cm}\overline\Lambda
     \overline\Lambda,\overline\Theta\overline\Theta\hspace{2.65cm}
     -\frac{1}{3}e\hspace{1.65cm}(3,1,1)\hspace{2mm}(3,3,1)
     \hspace{2mm}(3,1,3).\hspace{5mm}                                 \label{7}
 \end{eqnarray}
 \setcounter{equation}{\value{enumi}}}
 In this step the confining forces are, in kind, in $SU(3)\otimes
 {SU(2)}_L\otimes{SU(2)}_R$ gauge symmetry but the 
 $SU(2)_L\otimes{SU(2)}_R$ confining forces must be main 
 because of the energy scale of $\Lambda_L,\Lambda_R>>\Lambda_c$
 and then the color gauge coupling $\alpha_s$ and e.m. coupling 
 constant $\alpha$ are negligible. As is well known,
 the coupling constant of $SU(2)$ confining force are 
 characterized 
 by $\varepsilon_i=\sum_a\sigma_p^a\sigma_q^a$,where 
 ${\sigma}s$ are $2\times2$ matrices of $SU(2)$, $a=1,2,3$, 
 $p,q=\Lambda,\overline\Lambda,\Theta,\overline
 \Theta$, $i=0$ for singlet and $i=3$ for triplet. 
 They are calculated as 
 $\varepsilon_0=-3/4$ which causes the attractive force and 
 and $\varepsilon_3=1/4$ causing the repulsive force. Next, 
 $SU(3)_C$ octet and sextet states are repulsive but singlet,
 triplet and antitriplet states are attractive and then the formers 
 are disregarded. Like this, two primons are confined into composite 
 objects in more than  one singlet state of any 
 $SU(3)_C,SU(2)_L,SU(2)_R$. Note that three primon systems 
 cannot make the singlet states of $SU(2)$. Then we omit them. 
 In Eq.(7b), the $(1,1,1)$-state is the ``most attractive channel''. 
 Therefore $(\Lambda\overline\Theta)$, $(\overline\Lambda
 \Theta)$, $(\Lambda\overline\Lambda)$ and $(\Theta\overline
 \Theta)$ of $(1,1,1)$-states with neutral e.m. charge must 
 have been most abundant in the universe. 
 Further $(\overline{3},1,1)$- and 
 $(3,1,1)$-states in Eq.(7a,c) are next attractive. 
 They presumably go into $\{(\Lambda\Theta)(\overline\Lambda
 \overline\Theta)\}, \{(\Lambda\Lambda)(\overline\Lambda 
 \overline\Lambda)\}$, etc. of $(1,1,1)$-states 
 with neutral e.m. charge.
 These objects may be the candidates for the ``cold dark matters'' 
 if they have even tiny masses. 
 It is presumable that the ratio of the quantities between 
 the ordinary matters and the dark matters 
 firstly depends on the color and hypercolor charges 
 and the quantity of the latter much exesses that of the former 
 (maybe the ratio is more than $1/(3\times3)$). 
 Finally the $(*,3,1)$-and $(*,1,3)$-states are remained 
 ($*$ is $1,3,\overline{3}$). 
 They are also stable
 because $|\varepsilon_0|>|\varepsilon_3|$. They are, so to say,  
 the ``intermediate clusters'' towards constructing 
 ordinary matters(quarks,leptons and ${\bf W,Z}$).
\footnote{Such thoughts have been proposed by Maki in Ref.[22]}
  Here we call such intermediate clusters ``subquarks'' 
  and denote them as follows :
\newline
\hspace*{6.5cm}$Y.M.representation$\hspace{1.5cm}$spin$
 \hspace{0.5cm}$e.m.charge$
{
 \setcounter{enumi}{\value{equation}}
 \addtocounter{enumi}{1}
 \setcounter{equation}{0} 
 \renewcommand{\theequation}{\theenumi\alph{equation}}
 \begin{eqnarray}
    {\bf \alpha}&=&(\Lambda\Theta)\hspace{2.2cm}{\bf \alpha}_L:
    (\overline{3},3,1)\hspace{3mm}{\bf \alpha}_R:(\overline{3},1,3)
    \hspace{1.2cm}\frac{1}{2}\hspace{1.2cm}\frac{1}{3}e\\
    {\bf \beta}&=&(\Lambda\overline\Theta)\hspace{2.2cm}{\bf \beta}_L: 
    (1,3,1)\hspace{3mm}{\bf \beta}_R:(1,1,3)\hspace{1.3cm}\frac{1}{2}
    \hspace{1.3cm}0\\
    {\bf x}&=&(\Lambda\Lambda,\hspace{2mm}
    \Theta\Theta)\hspace{1.2cm}{\bf x}_L:
    (\overline{3},3,1)\hspace{3mm}{\bf x}_R:(\overline{3},1,3)
    \hspace{1.3cm}0\hspace{1.3cm}\frac{1}{3}e\\
    {\bf y}&=&(\Lambda\overline\Lambda\hspace{2mm}\Theta\overline\Theta)
    \hspace{1.4cm}{\bf y}_L:(1,3,1)\hspace{3mm}{\bf y}_R:(1,1,3)
    \hspace{1.3cm}0\hspace{1.3cm}0,                                 \label{8}
 \end{eqnarray}
 \setcounter{equation}{\value{enumi}}}
and there are also their antisubquarks[9].
\footnote{The notations of ${\bf \alpha}$,${\bf \beta}$,
 ${\bf x}$ and ${\bf y}$ are inherited from those in Ref.[9] 
 written by Fritzsch and Mandelbaum, because ours is, 
 in the subquark level, similar to theirs
 with two fermions and two bosons.
 R. Barbieri, R. Mohapatra and A. Masiero proposed 
 the similar model[9].}
\par
 Now we come to the step to build quarks and leptons. The gauge 
 symmetry of the confining forces in this step is also 
 $SU(2)_L\otimes{SU(2)}_R$ because the subquarks are in the
 triplet states of $SU(2)_{L,R}$ and then they are combined 
 into singlet states by the decomposition of $3\times3=1+3+5$
 in $SU(2)$. We make the first generation of quarks and leptons 
 as follows :
\newline
 \hspace*{8cm}$e.m.charge$\hspace{1.2cm}$Y.M.representation$
 {
 \setcounter{enumi}{\value{equation}}
 \addtocounter{enumi}{1}
 \setcounter{equation}{0}
 \renewcommand{\theequation}{\theenumi\alph{equation}} 
 \begin{eqnarray}
       <{\bf u}_h|&=&<{\bf \alpha}_h{\bf x}_h|\hspace{3.4cm}
                     \frac{2}{3}e\hspace{2,95cm}(3,1,1)\\
        <{\bf d}_h|&=&<\overline{\bf \alpha}_h\overline{\bf x}_h
                     {\bf x}_h|\hspace{2.5cm}-\frac{1}{3}e
                     \hspace{2.9cm}(3,1,1)\\
        <{\bf \nu}_h|&=&<{\bf \alpha}_h\overline{\bf x}_h|
                     \hspace{3.5cm}0\hspace{3.2cm}(1,1,1)\\
        <{\bf e}_h|&=&<\overline{\bf \alpha}_h\overline{\bf x}_h
                     \overline{\bf x}_h|\hspace{2.6cm}-e
                     \hspace{3.2cm}(1,1,1),                      \label{9}
 \end{eqnarray}
 \setcounter{equation}{\value{enumi}}}
 where $h$ stands for $L$(left handed) or $R$(right handed)[5].
\footnote{Subquark configurations in Eq.(9) are essentially the 
 same as those in Ref.[5] written by Kr\' olikowski, 
 who proposed the model of one fermion and one boson 
 with the same e.m. charge $e/3$}.
 Here we note that ${\bf \beta}$ and ${\bf y}$ do not appear.  
 In practice ($({\bf \beta}{\bf y}):(1,1,1)$)-particle 
 is a candidate for neutrino. But as Bjorken has pointed
 out[3], non-vanishing charge radius of neutrino is necessary
 for obtaining the correct low-energy effective weak interaction
 Lagrangian. Therefore ${\bf \beta}$ is assumed not to 
 contribute to forming ordinary quarks and leptons.
 However $({\bf \beta}{\bf y})$-particle may be a candidate
 for ``sterile neutrino''. 
 Presumably composite (${\bf \beta}$${\bf \beta}$)-;
 (${\bf \beta}\overline{\bf \beta}$)-;($\overline{\bf \beta}
 \overline{\bf \beta}$)-states may go into the dark matters. 
 It is also noticeable that in this model the leptons have 
 finite color charge radius and then $SU(3)$ gluons interact 
 directly with the leptons at energies of the order of, or 
 larger than $\Lambda_{L}$ or $\Lambda_{R}$[19]. 
 \par
 Concerning the confinements of primons and subquarks, the  
 confining forces of two steps are in the same spontaneously 
 unbroken $SU(2)_L\otimes{SU(2)}_R$ gauge symmetry.
 It is known that the running coupling constant 
 of the $SU(2)$ gauge theory satisfies the following equation : 
 {
 \setcounter{enumi}{\value{equation}}
 \addtocounter{enumi}{1}
 \setcounter{equation}{0}
 \renewcommand{\theequation}{\theenumi\alph{equation}}
 \begin{eqnarray}
       \frac{1}{\alpha^{a}_{W}(Q_{1}^{2})}&=&
                \frac{1}{\alpha^{a}_{W}(Q_{2}^{2})}
                +b_{2}(a)\ln\left(\frac{Q_{1}^{2}}{Q_{2}^{2}}\right),\\
       b_{2}(a)&=&\frac{1}{4\pi}\left(\frac{22}{3}-\frac{2}{3}
       \cdot{N}_{f}-\frac{1}{12}\cdot{N}_{s}\right),                 \label{10}
 \end{eqnarray}
 \setcounter{equation}{\value{enumi}}}
 where $N_f$ and $N_s$ are the numbers of fermions and scalars   
 contributing to the vacuum polarizations,
 ($a=q$) for the confinement of subquarks in quark and ($a=sq$)
 for confinement of primons in subquark.
 We calculate $b_2(q)=0.35$ which comes from that the number of 
 confined fermionic subquarks are $4$ (${\bf \alpha}_{i}, i=1,2,3$
 for color freedom, ${\bf \beta}$) and $4$ for bosons (${\bf x}_i,
 {\bf y}$) contributing to the vacuum polarization, and 
 $b_2(sq)=0.41$ which  is calculated with three kinds of 
 $\Lambda$ and $\Theta$ owing to three color freedoms. 
 Experimentary it is reported that $\Lambda_q>1.8$ TeV(CDF exp.)[13]
 or $\Lambda_q>2.4$ TeV(D0 exp.)[12].
 Extrapolations of $\alpha^{q}_{W}$ and $\alpha^{sq}_{W}$ to near 
 Plank scale are expected to converge to the same point and then
  tentatively, setting 
 $\Lambda_q=5$ TeV, $\alpha^{q}_{W}(\Lambda_q)=\alpha^{sq}_{W}
 (\Lambda_{sq})=\infty$, we get $\Lambda_{sq}=10^3\Lambda_q$,   
\par
 Next let us see the higher generations. Harari and Seiberg 
 have stated that the orbital and radial excitations 
 seem to have the wrong energy scale ( order of 
 $\Lambda_{L,R}$) and then the most likely type of 
 excitations is the addition of preon-antipreon pairs[6,25].
 In our model the essence of generation is like ``$isotope$"
 in case of nucleus.
 Then using ${\bf y}_{L,R}$ in Eq.(8,d) 
 we construct them as follows :
 {  
 \setcounter{enumi}{\value{equation}}
 \addtocounter{enumi}{1} 
 \setcounter{equation}{0}
 \renewcommand{\theequation}{\theenumi\alph{equation}}
 \begin{eqnarray}
 &&\left\{
 \begin{array}{lcl}
          <{\bf c}|&=&<{\bf \alpha}{\bf x}{\bf y}|\\
          <{\bf s}|&=&<\overline{\bf \alpha}\overline{\bf x}
                      {\bf x}{\bf y}|,
 \end{array} 
 \right.
 \hspace{6mm}
 \left\{
 \begin{array}{lcl}
          <{\bf \nu_\mu}|&=&<{\bf \alpha}\overline{\bf x}{\bf y}|\\
          <{\bf \mu}\hspace{2mm}|&=&<\overline{\bf \alpha}\overline{\bf x}
                                    \overline{\bf x}{\bf y}|,
 \end{array}
 \right.
 \hspace{0.7cm}\mbox{2nd generation}\\
 &&\left\{
 \begin{array}{lcl}
          <{\bf t}|&=&<{\bf \alpha}{\bf x}{\bf y}{\bf y}|\\
          <{\bf b}|&=&<\overline{\bf \alpha}\overline{\bf x}
                      {\bf x}{\bf y}{\bf y}|,
 \end{array}
 \right.
 \hspace{0.3cm}
 \left\{
 \begin{array}{lcl}
          <{\bf \nu_\tau}|&=&<{\bf \alpha}\overline{\bf x}
                              {\bf y}{\bf y}|\\
          <{\bf \tau}\hspace{2mm}|&=&<\overline{\bf \alpha}
          \overline{\bf x}\overline{\bf x}{\bf y}{\bf y}|,
 \end{array}
 \right.
 \hspace{4mm}\mbox{3rd generation},                               \label{11}
 \end{eqnarray} 
 \setcounter{equation}{\value{enumi}}}
 where the suffix $L,R$s are omitted for brevity. 
 We can also make vector and scalar particles with (1,1,1) : 
 {
 \setcounter{enumi}{\value{equation}}
 \addtocounter{enumi}{1} 
 \setcounter{equation}{0}
 \renewcommand{\theequation}{\theenumi\alph{equation}}
 \begin{eqnarray}&&\left\{
 \begin{array}{lcl}
             <{\bf W}^+|&=&<{\bf \alpha}^\uparrow{\bf \alpha}^
                             \uparrow{\bf x}|\\
             <{\bf W}^-|&=&<\overline{\bf \alpha}^\uparrow
                           \overline{\bf \alpha}^\uparrow\overline{\bf x}|,
 \end{array}
 \right.\hspace{6mm}
 \left\{
 \begin{array}{lcl}
             <{\bf Z}_1^0|&=&<{\bf \alpha}^\uparrow\overline
                             {\bf \alpha}^\uparrow|\\
             <{\bf Z}_2^0|&=&<{\bf \alpha}^\uparrow\overline
                             {\bf \alpha}^\uparrow{\bf x}\overline{\bf x}|, 
 \end{array} 
 \right.\hspace{0.3cm}\mbox{Vector}\\
 &&\left\{
 \begin{array}{lcl}
             <\hspace{2mm}{\bf S}^+\hspace{0.5mm}|
                        &=&<{\bf \alpha}^\uparrow{\bf \alpha}^
                           \downarrow{\bf x}|\\
             <\hspace{2mm}{\bf S}^-\hspace{0.5mm}|
                        &=&<\overline{\bf \alpha}^
                          \uparrow\overline{\bf \alpha}^\downarrow{\bf x}|,
 \end{array}
 \right.
 \hspace{6mm}
 \left\{
 \begin{array}{lcl}
            <{\bf S}_1^0|&=&<{\bf \alpha}^\uparrow\overline
                             {\bf \alpha}^\downarrow|\\
            <{\bf S}_2^0|&=&<{\bf \alpha}^\uparrow\overline
                            {\bf \alpha}^\downarrow{\bf x}\overline{\bf x}|,
 \end{array}
 \right.
 \hspace{3mm}\mbox{Scalar},\hspace{9mm}                                       \label{12}
 \end{eqnarray}
 \setcounter{equation}{\value{enumi}}}
 where the suffix $L,R$s are omitted for brevity and $\uparrow, 
 \downarrow$ indicate $spin\hspace{1mm}up, spin\hspace{1mm}down$ states.
 They play the role of intermediate bosons same as ${\bf \pi}$, 
 ${\bf \rho}$ in the strong interactions. As Eq.(9) and Eq.(12) 
 contain only ${\bf \alpha}$ and ${\bf x}$ subquarks, we can 
 draw the ``$line\hspace{1mm}diagram$'' of weak interactions as seen 
 in Fig (1). Eq.(9d) shows that the electron is constructed 
 from antimatters only. We know, phenomenologically, that this 
 universe is mainly made of protons, electrons, neutrinos,
 antineutrinos and unknown dark matters. It is said that  
 the universe contains  almost the same number of protons 
 and electrons.
  Our model show that one proton has the configuration
 of $({\bf u}{\bf u}{\bf d}): (2{\bf \alpha}, \overline{\bf \alpha}, 
 3{\bf x}, \overline{\bf x})$; electron: $(\overline{\bf \alpha}, 
 2\overline{\bf x})$; neutrino: $({\bf \alpha}, \overline{\bf x})$; 
 antineutrino: $(\overline{\bf \alpha}, {\bf x})$ and the dark 
 matters are presumably constructed from the same amount of matters 
 and antimatters because of their neutral charges. Note that 
 proton is a mixture of matters and anti-matters and electrons
 is composed of anti-matters only. This may lead the thought 
 that ``the universe is the matter-antimatter-even object.'' 
 And then there exists a conception-leap between 
 ``proton-electron abundance'' and ``matter abundance'' 
 if our composite scenario is admitted 
 (as for the possible way to realize the proton-electron 
 excess universe, see Ref.[14]).
 This idea is different from the current thought that
 the universe is made of matters only. Then the question 
 about CP violation in the early universe does not occur.
\par
 Our composite model contains two steps, namely the first is 
 ``subquarks made of primons'' and the second is ``quarks and 
 leptons made of subquarks''. Here let us discuss about 
 the mass generation mechanism  of quarks and leptons as 
 composite objects. Our model has only one  kind of fermion 
 : $\Lambda$  and boson : $\Theta$. The first  step of 
 ``subquarks made of primons'' seems to have nothing 
 to do with 'tHooft's anomaly matching condition[23] 
 because there is no global symmetry with 
 $\Lambda$ and $\Theta$. Therefore from this line of thought 
 it is  impossible to say anything about that ${\bf \alpha}$, 
 ${\bf \beta}$,  ${\bf x}$ and ${\bf y}$ are massless or massive. 
 However, if it is the case that the neutral  
 (1,1,1)-states of primon-antiprimon composites (as is stated above) 
 construct the dark matters, the masses of them are presumably less 
 than the order of MeV from the phenomenological aspects of 
 astrophysics. In this connection it is ineresting that Kr\'olikowski
 has showed one possibility of constructing ``massless'' composite
 particles(fermion-fermion or fermion-boson pair) controled
 by relativistic two-body equations[26].
 Then we may assume that these subquarks are 
 massless or almost massless compared with $\Lambda_{L,R}$ 
 in practice, that is, utmost a few MeV. 
 In the second step, the arguments of 'tHooft's anomaly 
 matching condition are meaningful. The confining of subquarks 
 must occur at the energy scale of $\Lambda_{L,R}>>\Lambda_c$ 
 and then it is natural that $\alpha_s, \alpha \rightarrow0$ 
 and that the gauge symmetry group is the spontaneously 
 unbroken $SU(2)_L\otimes{SU(2)}_R$ gauge group. 
 Seeing Eq.(9), we find quarks and leptons are composed of 
 the mixtures of subquarks and antisubquarks. 
 Therefore it is proper to 
 regard subquarks and antisubquarks as different kinds of 
 particles. From Eq.(8,a,b) we find eight kinds of fermionic 
 subquarks ( 3 for ${\bf \alpha}$, $\overline{\bf \alpha}$ and 
 1 for ${\bf \beta}$, $\overline{\bf \beta}$). So the global 
 symmetry concerned is $SU(8)_L\otimes{SU(8)}_R$. Then we  
 arrange :
 \begin{equation}
        ({\bf \beta},\overline{\bf \beta},{\bf \alpha}_i,\overline
        {\bf \alpha}_i\hspace{3mm}i=1,2,3\hspace{1mm})_{L,R}
        \hspace{2cm}in\hspace{1cm}(SU(8)_L\otimes{SU(8)}_R)_{global},                                                                           \label{13}
 \end{equation}
 where $i$ is color freedom.
 Next, the fermions in Eq.(13) are confined into the singlet 
 states of the local $SU(2)_L\otimes{SU(2)}_R$ gauge symmetry 
 and make up quarks and leptons as seen in Eq.(9) (eight fermions). 
 Then we arrange :
 \begin{equation}
        ({\bf \nu_e},{\bf e},{\bf u}_i,{\bf d}_i\hspace{3mm}i=1,2,3
        \hspace{1mm})_{L,R}\hspace{2cm}in\hspace{1cm}(SU(8)_L\otimes
        {SU(8)}_R)_{global},                                    \label{14}
 \end{equation}
 where $i$s are color freedoms. From Eq.(13) and Eq.(14) 
 the anomalies of the subquark level and the quark-lepton level 
 are matched and then all composite quarks and leptons (in the 1st 
 generation) are remained massless or almost massless. 
 Note again that presumably, 
 ${\bf \beta}$ and $\overline{\bf \beta}$ in Eq.(13) are composed 
 into ``bosonic'' (${\bf \beta}$${\bf \beta}$), 
 (${\bf \beta}$$\overline{\bf \beta}$) and 
 ($\overline{\bf \beta}$$\overline{\bf \beta}$), 
 which vapour out to the dark matters. Schrempp and Schrempp have 
 discussed about a confining $SU(2)_L\otimes{SU(2)}_R$ gauge 
 model with three fermionic preons and stated that it is 
 possible that not only the left-handed quarks and leptons   
 are composite but also the right-handed are so on the condition  
 that $\Lambda_R/\Lambda_L$ is at least of the order of $3$[11].
  As seen in Eq.(12a) the existence of composite 
 ${\bf W}_R$, ${\bf Z}_R$ is predicted. As concerning, 
 the fact that they are not observed yet means  that the masses of 
 ${\bf W}_R$, ${\bf Z}_R$ are larger than those of 
 ${\bf W}_L$, ${\bf Z}_L$ because of $\Lambda_R>\Lambda_L$. 
 Owing to 'tHooft's 
 anomaly matching condition the small mass nature of the 1st 
 generation comparing to $\Lambda_L$ is guaranteed but 
 the evidence that the quark masses of the 2nd and the 3rd 
 generations become larger as the generation numbers increase 
 seems to have nothing to do 
 with the anomaly matching mechanism in our model, because, as 
 seen in Eq.(11a,b), these generations are obtained by just 
 adding neutral scalar ${\bf y}$-particles. 
  This is different from Abott and Farhi's model 
 in which all fermions of three generations are equally
  embedded in $SU(12)$ global symmetry group and all 
  members take part in the anomaly matching mechanism[8]. 
  Concerning this, let us discuss a little about subquark 
  dynamics inside quarks. 
 According to ``Uncertainty Principle'' the radius of 
 the composite particle is, in general, roughly inverse 
 proportional to the kinetic energy of the constituent 
 particles moving inside it. 
 The radii of quarks may be around $1/\Lambda_{L,R}$ .
 So the kinetic energies of subquarks may be more than 
 hundreds GeV and then it is considered that the masses of 
 quarks essentially depend on the kinetic energies of 
 subquarks and such a large binding energy as 
 counterbalances them. As seen in Eq.(11a,b) our model 
 shows that the more the generation number increases 
 the more the number of the constituent particles 
 increases. So assuming that the radii 
 of all quarks do not vary so much (because we have no 
 experimental evidences yet), the interaction length among 
 subquarks inside quarks becomes shorter as generation 
 numbers increase and accordingly the average kinetic 
 energy per one subquark may increase. 
 Therefore integrating out the details of subquark 
 dynamics it could be said that 
 the feature of increasing masses of the 2nd and 
 the 3rd generations is essentially 
 described as a increasing function of the sum of 
 the kinetic energies of constituent subquarks. 
 From Review of Particle Physics[24] 
 we can phenomenologically parameterized the mass 
 spectrum of quarks and leptons as follows : 
 {
 \setcounter{enumi}{\value{equation}}
 \addtocounter{enumi}{1}
 \setcounter{equation}{0}
 \renewcommand{\theequation}{\theenumi\alph{equation}}
 \begin{eqnarray}
         M_{UQ}&=&1.2\times10^{-4}\times(10^{2.05})^n\hspace{1cm}
                  \mbox{GeV}\hspace{1.5cm}\mbox{for}\hspace{2mm}
                  \mbox{{\bf u},{\bf c},{\bf t}},\\
         M_{DQ}&=&3.0\times10^{-4}\times(10^{1.39})^n\hspace{1cm}
                 \mbox{GeV}\hspace{1.5cm}\mbox{for}\hspace{2mm}
                 \mbox{{\bf d},{\bf s},{\bf b}},\\
         M_{DL}&=&3.6\times10^{-4}\times(10^{1.23})^n\hspace{1cm}
                 \mbox{GeV}\hspace{1.5cm}\mbox{for}\hspace{2mm}
                 \mbox{{\bf e},${\bf \mu}$,${\bf \tau}$},      \label{15}
 \end{eqnarray}
 \setcounter{equation}{\value{enumi}}}
 where $n=1,2,3$ are the generation numbers and input data 
 are quark masses of 2nd and 3rd generation. 
 They seem to be  geometricratio-like. 
 The slope parameters of the up-quark 
 sector and down-quark sector are different, 
 so it is likely  that each has different aspects 
 in subquark dynamics. 
 It is interesting that the slope parameters of both down 
 sectors of quark and lepton are almost equal, which suggests 
 that there exist similar properties in substructure dynamics 
 and if it is the case, the slope parameter of up-leptonic(
 neutrino) sector may be the same as that of up-quark sector,
 that is, $M_{UL}\sim10^{2n}$.
 From Eq.(15) we obtain $M_{\bf u}=13.6$ MeV, 
 $M_{\bf d}=7.36$ MeV and $M_{\bf e}=6.15$ MeV. 
 These are a little unrealistic compared with the 
 experiments[24]. But considering the above discussions 
 about the anomaly matching conditions (
 Eq.(13,14)), it is natural that the masses of the 
 members of the 1st generation are roughly equal to 
 those of the subquarks, that is, a few MeV. 
 The details of their real mass-values may also depend on 
 the subquark dynamics owing to the effects of 
 electromagnetic and color gauge interactions. 
 These mechanism has studied by Weinberg[29] 
 and Fritzsch[30].
\par
 One of the experimental evidences inspiring the SM is 
 the ``universality'' of the coupling strength among the 
 weak interactions. Of course if the intermediate bosons are 
 gauge fields, they couple to the matter fields universally. 
 But the inverse of this statement is not always true, namely 
 the quantitative equality of the coupling strength of the 
 interactions does not necessarily imply that the intermediate 
 bosons are elementary gauge bosons. In practice the interactions 
 of ${\bf \rho}$ and ${\bf \omega}$ are regarded as indirect 
 manifestations of QCD. In case of chiral 
 $SU(2)\otimes{SU(2)}$ the pole dominance works very well
 and the predictions of current algebra and PCAC seem  
 to be fulfilled within about $5$\%[19]. 
 Fritzsch and Mandelbaum[9,19] and Gounaris, K\"ogerler 
 and Schildknecht[10,27] have elaborately discussed about 
 universality of weak interactions appearing as a 
 consequence of current algebra and ${\bf W}$-pole 
 dominance of the weak spectral functions from the 
 stand point of the composite model. 
 Extracting the essential points from their  
 arguments we mention our case as follows. 
 In the first generation let the weak charged currents be 
 written in terms of the subquark fields as :
 \begin{equation}
           {\bf J}_{\mu}^{+}=\overline{U}h_{\mu}D,\hspace{2cm}
            {\bf J}_{\mu}^{-}=\overline{D}h_{\mu}U,             \label{16}
 \end{equation}
 where $U=({\bf \alpha}{\bf x})$, $D=(\overline{\bf \alpha}
 \overline{\bf x}{\bf x})$ and $h_{\mu}=\gamma_{\mu}
 (1-\gamma_5)$.
 Reasonableness of Eq.(16) may given by the fact that
 $M_W<<\Lambda_{L,R}$ (where $M_W$ is ${\bf W}$-boson mass).
 Further, let $U$ and $D$ belong to the doublet of 
 the global weak isospin $SU(2)$ group and ${\bf W}^+$, 
 ${\bf W}^-$, $(1/\sqrt{2})({\bf Z}_1^0-{\bf Z}_2^0)$ 
 be in the triplet and $(1/\sqrt{2})({\bf Z}_1^0+
 {\bf Z}_2^0)$ be in the singlet of $SU(2)$. 
 These descriptions seem to be natural 
 if we refer the diagrams in Fig.(1). 
 The universality of the weak interactions are inherited 
 from the universal coupling strength of the algebra 
 of the global weak isospin $SU(2)$ group with the 
 assumption of ${\bf W}$-, ${\bf Z}$-pole dominance. 
 
   \section{The mass of $Z^{0}$'s scalar-partner}
      \hspace*{\parindent}
 
Eq.(12a,b) shows that the difference in $Z^{0}$ 
and $S^{0}$ essentially originates from the combination of 
two spins(up-spin and down-spin) of ${\bf \alpha}$- and
$\overline{\bf \alpha}$- subquark. $S^{0}$ has the 
combination of up- and down-spin and $Z^{0}$ has that of 
up- and up-spin.This situation is similar 
to hadronic mesons. They are the composite objects of 
 a quark($q$) and a anti-quark($\overline{q}$). namely, 
${\rho}$-${\pi}$, $K^{*}$-$K$, $D^{*}$-$D$, $B^{*}$-$B$.
Each vector meson mass(denoted by $M(V)$) is larger than 
the mass(denoted by $M(Ps)$) of its  pseudo-scalar partner. 
The mass differences between $M(V)$ and $M(Ps)$ are 
qualitatively explained by the hyperfine spin-spin 
interaction in Breit-Fermi Hamiltonian[28].
As the model of the hadronic mass spectra by the 
Breit-Fermi Hamiltonian is described by use of the 
semi-relativistical approach, it has some defects 
in the quantitative estimations, especially in the 
small mass mesons (such as ${\rho}$-${\pi}$ and 
$K^{*}$-$K$) but qualitatively it is not so bad, 
namely the explanation of the fact that : $M(V)>
M(Ps)$(and else $M(J=3/2\hspace{1.5mm}
 \rm{baryon})>M(J=1/2\hspace{1.5mm} \rm{baryon})$).
The hyperfine interaction Hamiltonian(denoted 
by $H_{q{\overline q}}^{l}$) causing mass split between 
$M(V)$ and $M(Ps)$ is described as : 

\begin{equation}
 H_{q{\overline{q}}}^{l=0}=
         -\frac{8{\pi}}{3m_{q}m_{\overline{q}}}
         \overrightarrow{S}_{q}\overrightarrow{S}_
         {\overline{q}}\delta(|\overrightarrow{r}|),       \label{17}
\end{equation} 
where $\overrightarrow{S}_{q(\overline{q})}$ is a 
operator of $q(\overline{q})$'s spin with its 
eigenvalue of 1/2 or -1/2, $m_{{q}(\overline{q})}$ is
quark (anti-quark) mass, $l$ is the orbital angular
momentum between $q$ and $\overline{q}$ and 
$|\overrightarrow{r}|=|\overrightarrow{r}_{q}
-\overrightarrow{r}_
{\overline{q}}|$[28]. 
\par
In QCD theory eight gluons are intermediate gauge 
bosons belonging to {\bf 8} representation which 
is real adjoint representation. Quarks(anti-quarks)
belong to ${\bf 3}({\bf \overline{3}})$ representation
which is complex fundamental representation. Therefore  
gluons can discriminate between quarks and anti-quarks 
and couple to them in the ''$different\hspace{2mm}sign$''.
\clearpage

The strength of their couplings to different color 
quarks and anti-quarks is described as :
 \begin{eqnarray}
    g\frac{\lambda_{ij}^{a}}{2}&:&\hspace{2.8cm}
    \rm{for\hspace{5mm} quarks}\nonumber\\
    -g\frac{\lambda_{ij}^{a}}{2}&:&\hspace{2.7cm}
    \rm{for\hspace{5mm} anti-quarks},                                                                                                                             \label{18}
 \end{eqnarray}
where $a(=1\sim8)$ : gluon indices; $i,j(=1,2,3)$ : 
quark indices; ${\lambda}$'s : SU(3) matrices
and $g$ : the coupling constant of gluons to quarks 
and anti-quarks(See Fig.(2)). 
The wave function of a color singlet 
$q\overline{q}$(meson) system is ${\delta}_{ij}/
\sqrt{3}$, corresponding to 
   $|q\overline{q}>=(1/\sqrt{3})\displaystyle
   {\sum_{i=1}^{3}}|q_{i}\overline{q_{i}}>$. 
By use of Eq.(18) the effective coupling for the 
$q{\overline{q}}$ system(denoted by ${\alpha}_{s}$)
is given by : 
 \begin{eqnarray}
     {\alpha}_{s}&=& {\displaystyle{\sum_{a,b}}{\sum_{i,j,k,l}}}
                      {\frac{1}{\sqrt{3}}}{\delta}_{ij}
                      \left({\frac{g}{2}}{\lambda}_{ik}^{a}\right)
                      \left(-{\frac{g}{2}}{\lambda}_{lj}^{b}\right)
                      {\frac{1}{\sqrt{3}}}{\delta}_{kl}
                      =-{\frac{g^{2}}{12}}{\displaystyle{\sum_{a,b}}{\sum_{j,l}}}
                      {\lambda}_{jl}^{a}{\lambda}_{lj}^{b}\nonumber\\
                 &=&-{\frac{g^{2}}{12}}\displaystyle{\sum_{a,b}}
                      \rm{Tr}\left({\lambda}^{a}{\lambda}^{b}\right)
                      =-{\frac{g^{2}}{6}}{\displaystyle
                      {\sum_{ab}}}{\delta}_{ab}\nonumber\\
                 &=&-{\frac{4}{3}}g^{2}.
                                                                   \label{19}
 \end{eqnarray}
 
Making use of Eq.(17) and Eq.(19) let us write the quasi-static 
Hamiltonian for a bound state of a quark and a anti-quark is
given as :
 \begin{equation}
     H=H_{0}+{\alpha}_{s}H_{q{\overline{q}}}^{l=0}.                      \label{20}   
 \end{equation}
Calculating the eigenvalue of $H$ in Eq.(20) we have :

 \begin{equation}
      M(V{\rm{or}}S)=M_{0}+{\xi}_{q}
      <\overrightarrow{S}_{q}\overrightarrow{S}_{\overline{q}}>, \label{21}
 \end{equation}
 where ${\xi}_{q}$ is a positive constant which incldes 
 the calculation of ${\alpha}_{s}$.
 In Eq.(21) it is found that  
 $<\overrightarrow{S}_{q}\overrightarrow{S}_{\overline{q}}>=-3/4$
 for pseudoscalar mesons and 
 $<\overrightarrow{S}_{q}\overrightarrow{S}_{\overline{q}}>=1/4$
 for vector mesons and then we have :
  \begin{eqnarray}
    M(Ps)&=&M_{0}-{\frac{3}{4}}{\xi}_{q}\nonumber\\
    M(\hspace{1mm}V\hspace{1mm})\hspace{0.7mm}
    &=&M_{0}+{\frac{1}{4}}{\xi}_{q}.                         \label{22}
  \end{eqnarray}
 Eq.(22) results that : 
  \begin{equation}
     M(V)>M(Ps).                                                 \label{23}
  \end{equation}                                             
 \par
 Here let us turn discussons to ``intermediate weak bosons''.
 As seen in Eq.(12a,b) $Z^{0}$ weak boson has its scalar 
 partner $S^{0}$ and both of them contain ``$fermionic$''
 ${\alpha}_{L}$ and $\overline{\alpha}_{L}$ as subquark elements.  
 Referring Eq.(8a) we find that both of ${\alpha}_{L}$ and 
 $\overline{\alpha}_{L}$ belong to ``adjoint ${\bf 3}$'' state 
 of $SU(2)_{L}$(which is the real representation) 
  and then $SU(2)_{L}$-hypercolor gluons cannot distingush    
 ${\alpha}_{L}$ from $\overline{\alpha}_{L}$. Therefore 
 the hypercolor gluons couple to ${\alpha}_{L}$ and
  $\overline{\alpha}_{L}$ in the ``$same\hspace{2mm} sign$''.
  This point is distinguishably dfferent from hadronic mesons
  (Refer Eq.(18)).
 The wave function of a hypercolor singlet 
 (${\alpha}{\overline{\alpha}}$)-system is 
 ${\delta}_{ij}/{\sqrt{3}}$, corresponding to 
  $|{\bf \alpha}_{i}\overline{\bf \alpha}_{i}>
  =(1/{\sqrt{3}}){\displaystyle|{\sum_{i=1}^{3}}}|
 {\bf \alpha}_{i}\overline{\bf \alpha}_{i}>$
 where $i=1,2,3$ are different three states of the triplet 
 of $SU(2)_{L}$.
 
  The strength of their couplings to 
 diferrent hypercolor subquarks and anti-subquarks 
 is described as :
 \begin{eqnarray}
    g_{h}\frac{{\tau}_{ij}^{a}}{2}
              &:&\hspace{2.5cm}
              \rm{for\hspace{6mm}subquark}\nonumber\\
    g_{h}\frac{{\tau}_{ij}^{a}}{2}
              &:&\hspace{2.5cm}
              \rm{for\hspace{2mm}anti-subquark},               \label{24}
 \end{eqnarray}   
 where $a(=1,2,3)$ : hypercolor gluon indices; $i,j(=1,2,3)$ :
 subquark and anti-subquark indices and ${\tau}$ : $SU(2)$ matrices
 and $g_{h}$ : the coupling constant of hypergluons to the subquarks 
 and anti-subquarks(See Fig.(2)). 
 By use of Eq.(24) the effective coupling 
 (denoted by ${\alpha_{W}}$) is given by :
 
  \begin{eqnarray}
     {\alpha}_{W}&=& {\displaystyle{\sum_{a,b}}{\sum_{i,j,k,l}}}
                      {\frac{1}{\sqrt{3}}}{\delta}_{ij}
                      \left({\frac{g_{h}}{2}}{\tau}_{ik}^{a}\right)
                      \left({\frac{g_{h}}{2}}{\tau}_{lj}^{b}\right)
                      {\frac{1}{\sqrt{3}}}{\delta}_{kl}
                      ={\frac{g^{2}_{h}}{12}}{\displaystyle
                      {\sum_{a,b}}{\sum_{j,l}}}
                      {\tau}_{jl}^{a}{\tau}_{lj}^{b}\nonumber\\
                 &=&{\frac{g^{2}_{h}}{12}}\displaystyle{\sum_{a,b}}
                      \rm{Tr}\left({\tau}^{a}{\tau}^{b}\right)
                      ={\frac{g^{2}_{h}}{6}}{\displaystyle
                      {\sum_{ab}}}{\delta}_{ab}\nonumber\\
                 &=&{\frac{1}{2}}g^{2}_{h},
                                                                   \label{25}
 \end{eqnarray}
 where $a,b=1,2,3$; $i,j,k,l=1,2,3$. Note that ${\alpha}_{s}$ 
 (in Eq.(19)) is ``$negative$'' and ${\alpha}_{W}$(in Eq.(25)) 
 ``$positive$''. Through the same procedure as  hadronic mesons 
 the masses of $Z^{0}$ and $S^{0}$ are described as 
 :
 \begin{equation} 
    M(Z^{0}\hspace{0.6mm}\rm{or}\hspace{0.6mm}S^{0})
        =M_{0}-{\xi}_{sq}<\overrightarrow{S}_{\alpha}
        \overrightarrow{S}_{\overline{\alpha}}>,                 \label{26}
 \end{equation}
 where ${\xi}_{sq}$ is a positive constant which includes 
 the calculation of ${\alpha}_{W}$ and 
 $\overrightarrow{S}_{\alpha(\overline{\alpha})}$ is 
 the spin operator of ${\alpha}(\overline{\alpha})$.          
  In Eq.(26) it is calculated that 
 $<\overrightarrow{S}_{\alpha}\overrightarrow{S}_
 {\overline{\alpha}}>=-3/4$
 for scalar : $S^{0}$ and 
 $<\overrightarrow{S}_{\alpha}\overrightarrow{S}_
 {\overline{\alpha}}>=1/4$
 for vector : $Z^{0}$ and then we get :
  \begin{eqnarray}
    M(S^{0})&=&M_{0}+{\frac{3}{4}}{\xi}_{sq}\nonumber\\
    M(Z^{0})&=&M_{0}-{\frac{1}{4}}{\xi}_{sq}.                     \label{27}
  \end{eqnarray}
From this it follows that : 
  \begin{equation}
     M(S^{0})>M(Z^{0}).                                           \label{28}
  \end{equation}
 Here let us define :
  \begin{eqnarray}
  \tilde{M}
      &=&\frac{1}{2}\left(M(S^{0})+M(Z^{0})\right),\nonumber\\
  {\Delta}
      &=&M(S^{0})-M(Z^{0}),\nonumber\\
     R&=&\frac{\Delta}{\tilde{M}}.                                \label{29} 
 \end{eqnarray}
 Experimentally it is reported : $M(Z^{0})=91$GeV[24], 
 with which using Eq.(29) we obtain :
  \begin{eqnarray}
    R&=&
      0.1\hspace{1cm}M(S^{0})\approx{100}
      \hspace{2mm}\rm{GeV},\nonumber\\
    R&=&
      0.2\hspace{1cm}M(S^{0})\approx{110}
      \hspace{2mm}\rm{GeV}.
  \end{eqnarray}                                                 \label{30}
 Therefore if the existence of the scalar particle 
 whose mass is a little above $Z^{0}$'s mass
  is confirmed in future it may be a scalar partner of 
  $Z^{0}$ and that might suggest the possibility of the 
  subquark structure.

{\bf Acknowledgements}
\par
 We would like to thank Y. Matsui 
 and A.Takamura for useful discussions and 
 the hospitality at the laboratory of the 
 elementary particle physics in Nagoya University.


\clearpage
  

\setlength{\unitlength}{0.7mm}
\begin{figure}
\begin{center}
\begin{picture}(180,270)(15,-20)

\put(47,216){\makebox(7,7){${\bf \overline{\alpha}}$}}
\put(43,207){\makebox(7,7){${\bf \overline{x}}$}}
\put(35,213){\makebox(7,7){${\bf \overline u}$}}
\put(49,170){\makebox(7,7){${\bf \overline{\alpha}}$}}
\put(45,176){\makebox(7,7){${\bf \overline x}$}}
\put(42,182){\makebox(7,7){${\bf x}$}}
\put(35,173){\makebox(7,7){${\bf d}$}}
\put(132,221){\makebox(7,7)[bl]{${\bf \overline{\alpha}}$}}
\put(135,215){\makebox(7,7)[bl]{${\bf \overline x}$}}
\put(138,209){\makebox(7,7)[bl]{${\bf \overline x}$}}
\put(145,215){\makebox(7,7){${\bf e^-}$}}
\put(140,180){\makebox(7,7){${\bf x}$}}
\put(135,171){\makebox(7,7){${\bf \overline{\alpha}}$}}
\put(145,173){\makebox(7,7){${\bf \overline{\nu}}$}}
\put(90,207){\makebox(10,10)[b]{${\bf W^-}$}}
\put(83,205){\line(-2,1){28}}
\put(75,197){\line(-2,1){24}}
\put(75,197){\line(-2,-1){23}}
\put(86,197){\line(-2,-1){32}}
\put(83,190){\line(-2,-1){27}}
\put(86,185){\vector(1,0){15}}
\put(83,205){\line(1,0){22}}
\put(86,197){\line(1,0){14}}
\put(83,190){\line(1,0){22}}
\put(105,205){\line(2,1){25}}
\put(100,197){\line(2,1){32}}
\put(112,197){\line(2,1){22}}
\put(112,197){\line(2,-1){26}}
\put(105,190){\line(2,-1){29}}
\put(46,147){\makebox(7,7)[br]{${\bf {\alpha}}$}}
\put(42,141){\makebox(7,7)[br]{${\bf x}$}}
\put(38,135){\makebox(7,7)[br]{${\bf x}$}}
\put(39,110){\makebox(7,7){${\bf \overline{x}}$}}
\put(42,105){\makebox(7,7){${\bf \overline x}$}}
\put(46,100){\makebox(7,7){${\bf \overline{\alpha}}$}}
\put(135,147){\makebox(7,7)[bl]{${\bf \alpha}$}}
\put(139,142){\makebox(7,7)[bl]{${\bf \alpha}$}}
\put(143,135){\makebox(7,7)[bl]{${\bf x}$}}
\put(143,110){\makebox(7,7)[l]{${\bf \overline{x}}$}}
\put(139,105){\makebox(7,7)[l]{${\bf \overline{\alpha}}$}}
\put(135,100){\makebox(7,7)[l]{${\bf \overline{\alpha}}$}}
\put(35,140){\makebox(10,10){${\bf e^+}$}}
\put(35,100){\makebox(10,10){${\bf e^-}$}}
\put(145,140){\makebox(10,10){${\bf W^+}$}}
\put(145,100){\makebox(10,10){${\bf W^-}$}}
\put(88,129){\makebox(10,10){${\bf Z_1^0}$}}
\put(83,130){\line(-2,1){30}}
\put(80,125){\line(-2,1){30}}
\put(70,125){\line(-2,1){22}}
\put(70,125){\line(-2,-1){22}}
\put(80,125){\line(-2,-1){30}}
\put(86,115){\vector(1,0){15}}
\put(83,120){\line(-2,-1){30}}
\put(83,130){\line(1,0){20}}
\put(83,120){\line(1,0){20}}
\put(103,130){\line(2,1){30}}
\put(106,125){\line(2,1){30}}
\put(116,125){\line(2,1){22}}
\put(116,125){\line(2,-1){22}}
\put(106,125){\line(2,-1){30}}
\put(103,120){\line(2,-1){30}}
\put(46,87){\makebox(7,7)[br]{${\bf {\alpha}}$}}
\put(42,81){\makebox(7,7)[br]{${\bf x}$}}
\put(38,75){\makebox(7,7)[br]{${\bf x}$}}
\put(39,50){\makebox(7,7){${\bf \overline{x}}$}}
\put(42,45){\makebox(7,7){${\bf \overline x}$}}
\put(46,40){\makebox(7,7){${\bf \overline{\alpha}}$}}
\put(135,87){\makebox(7,7)[bl]{${\bf \alpha}$}}
\put(139,82){\makebox(7,7)[bl]{${\bf x}$}}
\put(143,75){\makebox(7,7)[bl]{${\bf \alpha}$}}
\put(143,50){\makebox(7,7)[l]{${\bf \overline{\alpha}}$}}
\put(139,45){\makebox(7,7)[l]{${\bf \overline{x}}$}}
\put(135,40){\makebox(7,7)[l]{${\bf \overline{\alpha}}$}}
\put(35,80){\makebox(10,10){${\bf e^+}$}}
\put(35,40){\makebox(10,10){${\bf e^-}$}}
\put(145,80){\makebox(10,10){${\bf W^+}$}}
\put(145,40){\makebox(10,10){${\bf W^-}$}}
\put(88,69){\makebox(10,10){${\bf Z_2^0}$}}
\put(83,70){\line(-2,1){30}}
\put(76,67){\line(-2,1){25}}
\put(70,65){\line(-2,1){22}}
\put(70,65){\line(-2,-1){22}}
\put(76,63){\line(-2,-1){25}}
\put(86,55){\vector(1,0){15}}
\put(83,60){\line(-2,-1){30}}
\put(83,70){\line(1,0){20}}
\put(76,67){\line(1,0){34.5}}
\put(76,63){\line(1,0){34.5}}
\put(83,60){\line(1,0){20}}
\put(103,70){\line(2,1){30}}
\put(110.5,67){\line(2,1){25}}
\put(116,65){\line(2,1){22}}
\put(116,65){\line(2,-1){22}}
\put(110.5,63){\line(2,-1){25}}
\put(103,60){\line(2,-1){30}}
\end{picture}
\end{center}
\caption{Subquark-line diagrams of the weak interactions}
\end{figure}
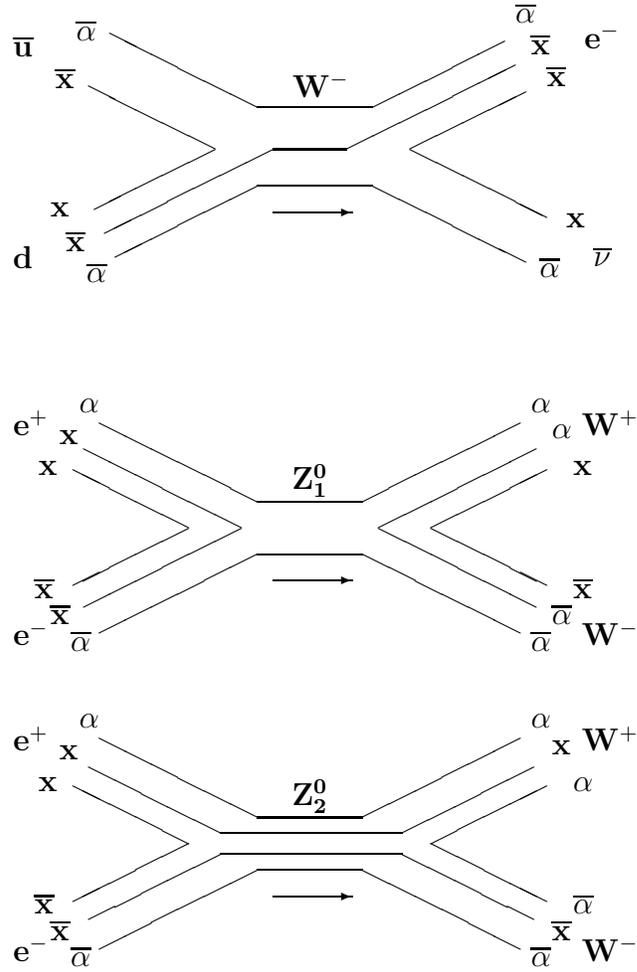

\setlength{\unitlength}{0.8mm}
\begin{figure}
\begin{center}
\begin{picture}(180,250)(15,-20)

\put(140,185){\makebox(30,30)[l]{(A)}}
\put(0,185){\makebox(30,30)[r]{${V}
             \hspace{1mm}({Ps})$}}
\put(60,220){\line(1,0){60}}
\multiput(90,185)(0,4){9}{\line(0,1){3}}
\put(66,221){\makebox(50,10)[b]{
            $+{g\displaystyle{\frac{\lambda_{ik}^{a}}{2}}}$}}
\put(50,216){\makebox(7,7)[r]{${\bf q}_{i}$}}
\put(123,216){\makebox(7,7)[l]{${\bf q}_{k}$}}
\put(50,181){\makebox(7,7)[r]{${\bf \overline{q}}_{j}$}}
\put(60,185){\line(1,0){60}}
\put(123,181){\makebox(7,7)[l]{${\bf \overline{q}}_{l}$}}
\put(66,173){\makebox(50,10)[b]{
            ${-g\displaystyle{\frac{\lambda_{jl}^{b}}{2}}}$}}

\put(140,95){\makebox(30,30)[l]{(B)}}
\put(0,95){\makebox(30,30)[r]{${Z^0}
             \hspace{1mm}({S^0})$}}
\put(60,130){\line(1,0){60}}
\multiput(90,95)(0,4){9}{\line(0,1){3}}
\put(66,131){\makebox(50,10)[b]{
            $+{g_{h}\displaystyle{\frac{\tau_{ik}^{a}}{2}}}$}}
\put(50,126){\makebox(7,7)[r]{${\bf \alpha}_{i}$}}
\put(123,126){\makebox(7,7)[l]{${\bf \alpha}_{k}$}}
\put(50,91){\makebox(7,7)[r]{${\bf \overline{\alpha}}_{j}$}}
\put(60,95){\line(1,0){60}}
\put(123,91){\makebox(7,7)[l]{${\bf \overline{\alpha}}_{l}$}}
\put(66,83){\makebox(50,10)[b]{
            $+{g_{h}\displaystyle{\frac{\tau_{jl}^{b}}{2}}}$}}

\end{picture}
\end{center}
\caption{
   (A) Gluon exchange in ${\bf q}\overline{\bf q}$ system;
    (B) Hypergluon exchange in
      ${\bf \alpha}\overline{\bf \alpha}$ system.
       }
\end{figure}

\end{document}